\documentstyle[a4,12pt,amssymb]{article}
\input epsf
\begin{document}
\baselineskip=20pt
{\center
{\bf \huge On the number of limit cycles of the Li\'enard equation\\}
\vspace{1 cm}
{\large H. Giacomini and S. Neukirch\\}
Laboratoire de Math\'ematiques
 et Physique Th\'eorique\\ C.N.R.S. UPRES A6083 \\
Facult\'e des Sciences et Techniques, Universit\'e de Tours\\
F-37200 Tours FRANCE\\
}
\vspace{3 cm}
{\center \section*{Abstract}}
{ \small
In this paper, we study a Li\'enard system of the form
 $\dot{x}=y-F(x) \;
, \; \dot{y}=-x$, where $F(x)$ is an odd polynomial.
 We introduce a method that gives a sequence of algebraic
 approximations to the equation of each limit cycle of the system.
This sequence seems to converge to the exact
 equation of each limit cycle.\\
We obtain also a sequence of polynomials $R_n(x)$ whose
 roots of odd multiplicity are related to the number and
 location of the limit cycles of the system.
}

\vspace{5 cm}

{\bf PACS numbers : 05.45.+b , 02.30.Hq , 02.60.Lj , 03.20.+i \\

Key words : Li\'enard equation, limit cycles.}
\clearpage
A two-dimensional autonomous dynamical system is defined
by two coupled first order differential equations of the form : \\
\begin{equation}
\dot{x}=P(x,y) \quad , \quad \dot{y}=Q(x,y) \label{ode}
\end{equation}
where $P$ and $Q$ are two functions of the variables $x$
and $y$ and the overdots denote a time derivative.

 Such a type
of dynamical system appears very often within several branches
of science, such as biology, chemistry, astrophysics, mechanics,
electronics, fluid mechanics, etc 
\cite{strogatz,nayfeh,kerhet,salamich2,poland,salamich}.

One of the most difficult
problems connected with the study of system (\ref{ode}) is the
question of the number of limit cycles. 
A limit cycle is an isolated closed trajectory. Isolated
 means that
the neighboring trajectories are not closed; they spiral
 either toward
or away from the limit cycle. If all neighboring
 trajectories aproach
the limit cycle, we say that the limit cycle is stable or
 attracting.
Otherwise the limit cycle is unstable or, in exeptional cases,
half-stable. Stable limit cycles are very important in science.
They model systems that exhibit self-sustained oscillations. In
other words, these systems oscillate even in the absence of 
external periodic forcing. Of the countless examples that could
 be given,
we mention only a few : the beating of a heart, chemical
 reactions that 
oscillate spontaneously, self-excited vibrations in bridges and
 airplane wings, etc. In each case, there is a standard
 oscillation of some preferred period, waveform and amplitude.
 If the system is slightly perturbated, it always returns to
 the standard cycle. Limit cycles are an inherently 
nonlinear phenomena; they cannot occur in linear systems
\cite{qian,maier,farkas,viano,llibre,delamotte}.

 The first physical model to appear in the 
literature which can be transformed to a system of type (\ref{ode})
containing a limit cycle is due to Rayleigh \cite{rayleigh}. The
following equation~:
\begin{equation}
\frac{d^2 y}{d t^2}+\epsilon \left( \frac{1}{3}(\frac{dy}{dt})^3
- \frac{dy}{dt} \right) + y=0 \label{equa rayleigh}
\end{equation}
\\ that originated in connection with a theory of the oscillation
of a violin string, was derived by Rayleigh in 1877.

 In 1927,
 the dutch scientist van der Pol \cite{van der pol} described 
self-excited oscillations in an electrical circuit with a triode
tube with resistive properties that change with the current. The
equation derived by van der Pol reads~: 
\begin{equation}
\frac{d^2 x}{d t^2} + \epsilon (x^2-1) \frac{d x}{d t} + x=0 \label{equa vdp}
\end{equation}
Equations (\ref{equa rayleigh}) and (\ref{equa vdp}) are equivalent, as
can be seen by differentiating (\ref{equa rayleigh}) with respect to $t$
and putting $\frac{d y}{d t}=x$.

 In 1928, the french engineer 
A. Li\'enard \cite{lienard} gave a criterion for the uniqueness
of periodic solutions for a general class of equations, for
which the van der Pol equation is a special case :
\begin{equation}
\frac{d^2 x}{d t^2}+f(x) \frac{d x}{d t} +x=0 \label{equa lienard}
\end{equation}
Li\'enard tranformed (\ref{equa lienard}) to a first order system by 
setting $\frac{d x}{d t}=z$, yielding
\begin{equation}
\frac{d x}{d t}=z \quad , \quad \frac{d z}{d t}=-x-f(x) z 
\label{equa lienard ode}
\end{equation}
In fact, in his proof, Li\'enard used a form equivalent to
 (\ref{equa lienard ode}), obtaining through the change of variable
 $z=y-F(x)$, where $F(x)=\int_0^x f(\tau) d\tau$ :
\begin{equation}
\frac{d x}{d t}=y-F(x) \quad , \quad \frac{d y}{d t}=-x 
\label{equa lienard ode bis}
\end{equation}
Equation (\ref{equa lienard}) is referred to as Li\'enard equation and
both system (\ref{equa lienard ode}) and (\ref{equa lienard ode bis}) are
called Li\'enard systems. They are a particular case of (\ref{ode}).

In 1942, Levinson and Smith \cite{smith} suggested the following 
generalization of system (\ref{equa lienard ode bis})~:
\begin{equation}
\frac{d x}{d t}=y-F(x) \quad , \quad \frac{d y}{d t}=-g(x) 
\label{equa lienard gene}
\end{equation}
or equivalently~:
\begin{equation}
\frac{d x}{d t}=z \quad , \quad \frac{d z}{d t}=-g(x)-f(x) z 
\label{equa lienard gene bis}
\end{equation}
Sytems (\ref{equa lienard gene}) and (\ref{equa lienard gene bis}) are 
equivalent to~:
\begin{equation}
\frac{d^2 x}{d t^2}+f(x) \frac{d x}{d t}+g(x)=0 \label{equa lienard fin}
\end{equation}
which is sometimes referred to as the generalized Li\'enard equation.\\

In this paper, we will consider the case $g(x)=x$ and $F(x)$ given by
an arbitrary odd polynomial of degree $m$. The fundamental problem for
this type of system is the determination of the number of limit
 cycles for a given polynomial $F(x)$ 
\cite{de melo,perko,lloyd,dumortier2,dumortier,blows and lloyd,xun,yang}~.
 For $m=3$, i.e. 
for $F(x)=a_1 x +a_3 x^3$, it has been shown in \cite{de melo} that
 the system
has a unique limit cycle if $a_1 a_3 <0$ and no limit cycle if $a_1 a_3 >0$.
For $m=5$ it has been shown in \cite{rychkov} that the 
maximum number of limit cycles
is two. For $m > 5$, there are no general results about the 
number of limit cycles of (\ref{equa lienard ode bis}).

In this paper, we present a new method that gives information about the
 number of 
limit cycles of (\ref{equa lienard ode bis}) and their location
in phase space, for a given odd polynomial F(x).
This method gives also a sequence of algebraic approximations to the cartesian 
equation of the limit cycles.

 We will explain our method through the
analysis of a very well known case, the van der Pol equation. In this
 case, we have~:\\ 
\begin{equation}
F(x)=\epsilon (x^3/3-x) \label{f de vdp}
\end{equation}
 We propose a function 
$h_2(x,y)=y^2+g_{1,2}(x) y+g_{0,2}(x)$, where $g_{1,2}(x)$
 and $g_{0,2}(x)$ are 
arbitrary functions of $x$.
Here, the second subindex makes reference to the degree of the polynomial
$h_2$ with respect to the $y$ variable.
 Then we calculate 
$\dot{h_2}=(y-F(x)) \frac{\partial h_2}{\partial x}-x 
\frac{\partial h_2}{\partial y}$. This quantity is a second degree
 polynomial in the variable $y$. We will choose $g_{1,2}(x)$ and
 $g_{0,2}(x)$ in such a way that the coefficients of
 $y^2$ and $y$ in $\dot{h_2}$ are zero. From these conditions
, we obtain $g_{1,2}(x)=k_1$ and $g_{0,2}(x)=x^2+k_0$, where $k_0$ and $k_1$ 
are arbitrary constants.
As $F(x)$ is an odd polynomial, if $(x,y)$ is a point of the limit cycle
of (\ref{equa lienard ode bis}), then the point $(-x,-y)$ also belongs to 
this limit cycle. The equation of a limit cycle of 
(\ref{equa lienard ode bis}) must be invariant by the transformation 
$(x,y) \rightarrow (-x,-y)$. We want the function $h_2(x,y)$ to have this
symmetry too. Thus we take $k_1=0$.
 We then have $\dot{h_2}=R_2(x)=-2 x F(x)=
-2 \epsilon x^2 (x^2/3-1)$. The polynomial $R_2(x)$ is even and it has 
exactly one positive root of odd multiplicity, i.e. $x=\sqrt{3}$.

 If we 
integrate the function $\dot{h_2}$ along the limit cycle, we have~:
$\int_0^T \dot{h_2}(x(t),y(t)) dt=\int_0^T R_2(x(t)) dt$
, where T is the period; but 
$\int_0^T \dot{h_2}(x(t),y(t)) dt=h_2(x(T),y(T))-h_2(x(0),y(0))=0$.
Consequently, we find~:
$\int_0^T R_2(x(t)) dt=0$.
This last equality tells us that there cannot be any limit cycle
in a region of the phase plane where $R_2(x)$ is of constant sign.
For the van der Pol system, $R_2(x)$ has a root of
odd multiplicity at $x=\sqrt{3}$
, hence the maximum value of $x$ for the limit cycle
 must be greater than $\sqrt{3}$.
The curves defined by $h_2(x,y)=x^2+y^2+k_0=0$ are closed for
$k_0<0$.

 As the next step of our procedure, we propose a fourth degree
polynomial in $y$ for the function $h_4(x,y)$, i.e.
$h_4(x,y)=y^4+g_{3,4}(x) y^3+g_{2,4}(x) y^2+g_{1,4}(x) y +g_{0,4}(x)$
(polynomials $h_n(x,y)$ with $n$ odd do not give useful information about
the limit cycles of the system since the level curves $h_n(x,y)$ 
are open and the polynomials $R_n(x)$ have always a single root of odd
multiplicity at $x=0$)
. By imposing the
condition that $\dot{h_4}$ must be a function of only $x$,
we find $\dot{h_4}=R_4(x)$, where $R_4(x)$ is an even
polynomial of tenth degree. The roots of $R_4(x)$ depend of $\epsilon$, hence
in the following, we will take $\epsilon=1$. For this case, $R_4(x)$ 
has only one
positive root of odd multiplicity, given by $x \simeq 1.824$. This 
root is greater 
than the root of $R_2(x)$. Obviously, the maximum value of $x$ for the limit
 cycle
must be greater than this value.

 We have in this way a new lower bound
for the maximum value of $x$ on the limit cycle. Moreover the number
 of positive roots 
of odd multiplicity is equal to the number of limit cycles of the system.
The condition that $\dot{h_4}$ must be a function only of $x$ imposes 
a first order
trivial differential equation for each function $g_{j,n}(x)$. These
equations can be solved by direct integration and we obtain in this way
all the functions $g_{j,n}(x)$. We take all the integration constants, that
appear when we solve these equations, equal to zero. In this way, the level
curves $h_4(x,y)=K$ are all closed for positive values of $K$
and even values of $n$. 
Moreover, the function $h_4(x,y)$ is a polynomial in $x$ and $y$.
 
We have found
the same results for greater values of $n$ even. We have calcultated $h_n(x,y)$
and $R_n(x)$ up to order 20. In all cases, the polynomials $R_n(x)$
have only one positive root of odd multiplicity. Let $r_n$ be the
 number of such roots.
 For the van der Pol equation, it seems that 
$r_n=1 \; \forall n$ even. These roots approach in a monotonous fashion the
maximum value of $x$ on the limit cycle. The functions $h_n(x,y)$
 are polynomials in $x$ and $y$ for all $n$. The level curves
 $h_n(x,y)=K$ are 
all closed for positive
values of K. By imposing the condition that the maximum value
 of $x$ on the curve $h_n(x,y)=K > 0$ must be equal to the root of
 $R_n(x)$, we find a particular value of K for each $n$ even. Let
 us call this value $K_n^\star$. The level curve $h_n(x,y)=K_n^\star$
 represents an algebraic approximation to the limit cycle.

\begin{figure}[p]
$$
\epsfxsize=5cm
\epsfbox{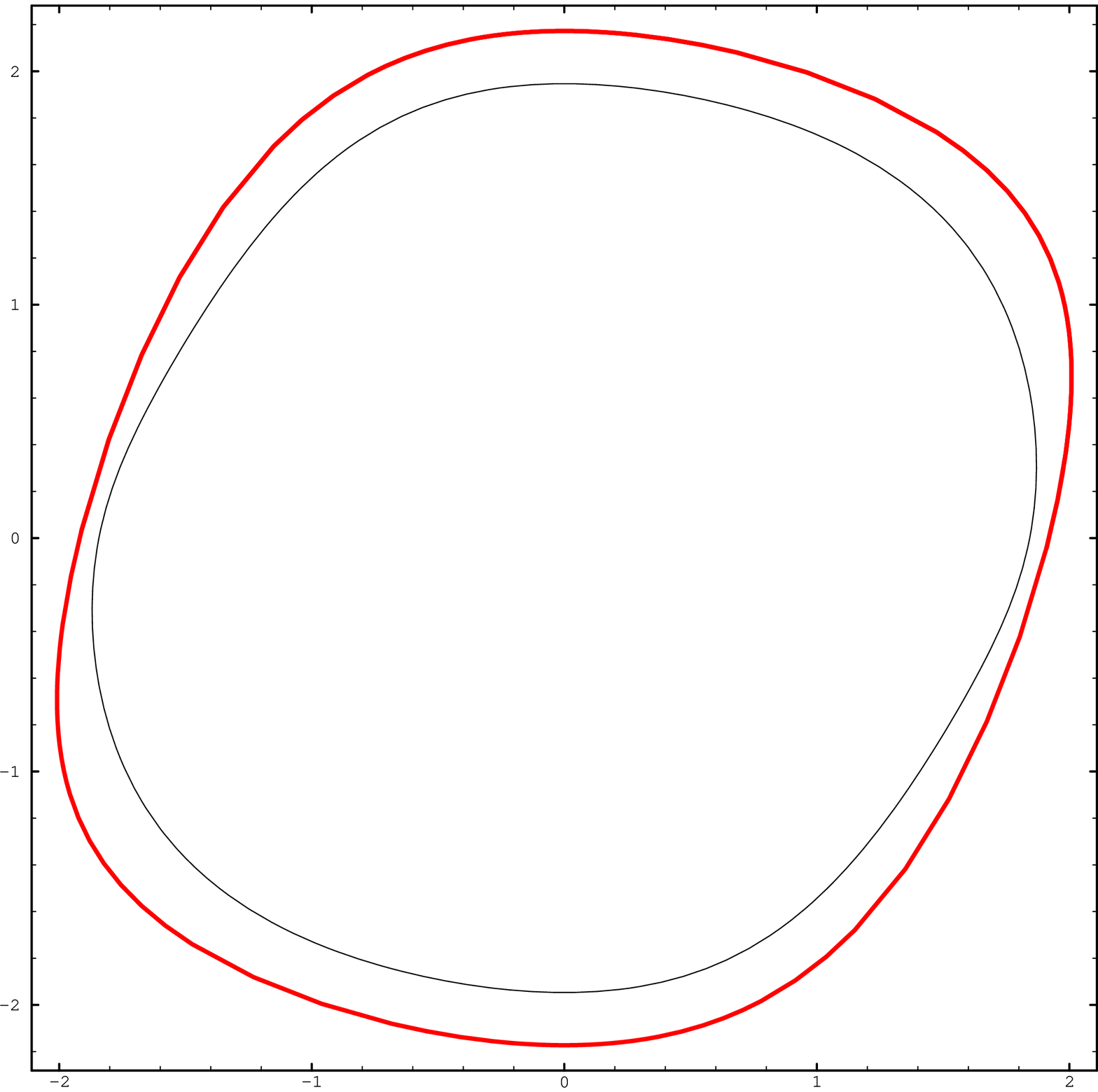}
$$
\caption{The limit cycle of the van der Pol equation (exterior curve)
and the algebraic approximation $h_6(x,y)=K_6 ^\star$.}
\label{fig vdp 6}
\end{figure}

\begin{figure}[p]
$$
\epsfxsize=5cm
\epsfbox{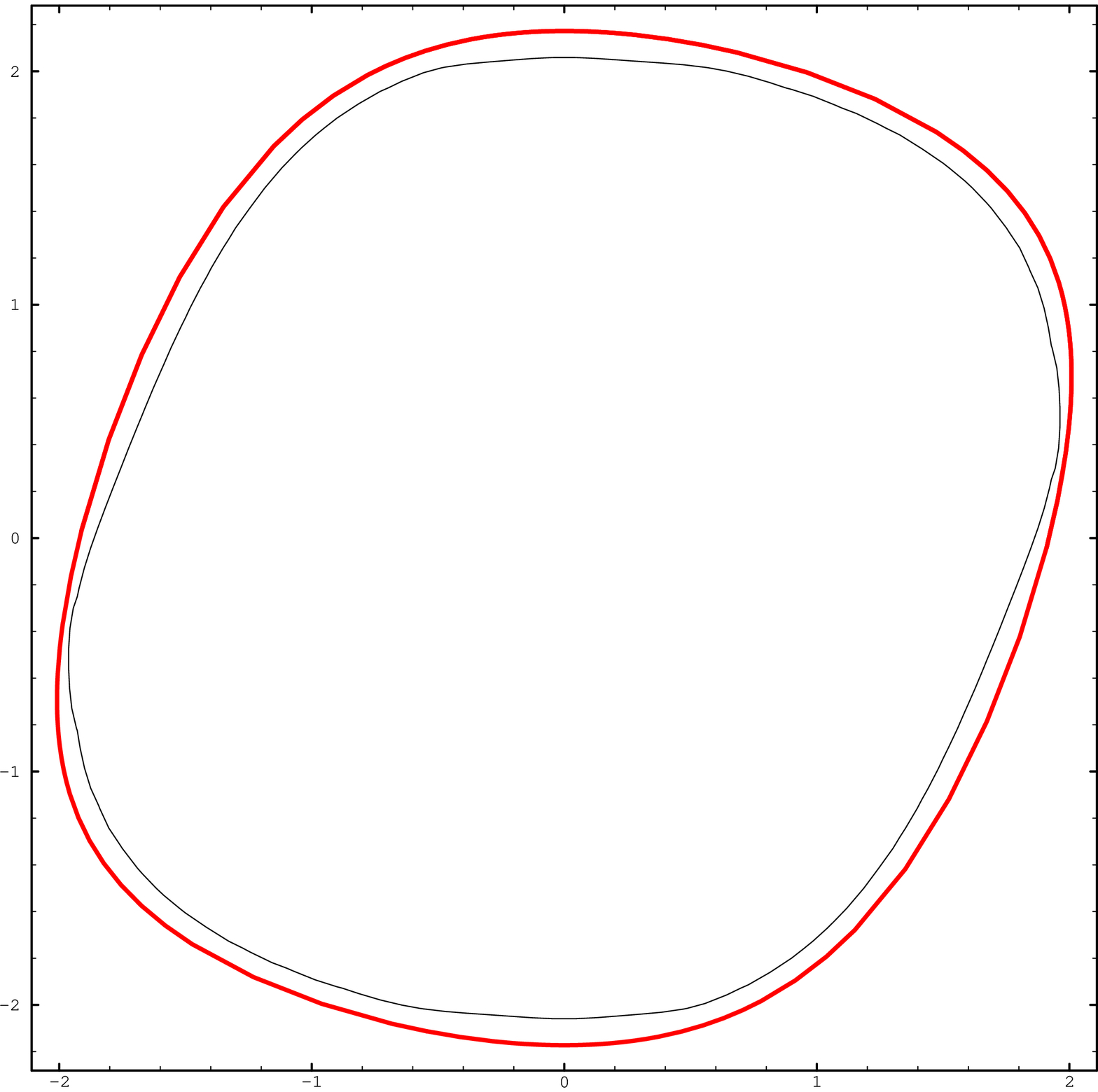}
$$
\caption{The limit cycle of the van der Pol equation (exterior curve)
and the algebraic approximation $ h_{18}(x,y)=K_{18} ^\star$ .}
\label{fig vdp 18}
\end{figure}

In fig. \ref{fig vdp 6} and \ref{fig vdp 18} we show this
 curve for the values $n=6$ and $n=18$, respectively.
In table \ref{table vdp} we give the values of the roots of $R_n(x)$
 and the values of
 $K_n ^\star $ for $2 \leq n \leq 20$.
 The numerical value of the
 maximum of $x$ on the limit cycle, determined from a numerical
 integration of (\ref{equa lienard ode bis}),
with $F(x)$ defined by (\ref{f de vdp}),
 is $x_{max} \simeq 2.01 \; (\epsilon =1)$.
It is clear that the roots of $R_n(x)$ seem to converge to $x_{max}$
 and the curves $f_n(x,y)=K_n ^\star $ seem to converge to the
 limit cycle.\\

We have also studied the case~:
\begin{equation}
F(x)=0.8 x-\frac{4}{3} x^3+0.32 x^5 \label{equa 2cy}
\end{equation}
This system has exactly two limit cycles \cite{perko}.
We have calculated the polynomials $h_n(x,y)$ and $R_n(x)$
up to $n=16$. The polynomials
 $R_n(x)$ have exactly two positive roots of odd multiplicity.
 We conjecture that $r_n=2 
\; \forall \, n$ even. For
 each value of $n$, we determine two values $K_{n1}^\star$ and 
$K_{n2} ^\star$. The closed curves $h_n(x,y)=K_{n1} ^\star $ 
and $h_n(x,y)= K_{n2} ^\star $ provide algebraic approximations to
 each cycle for each value of $n$ even.

\begin{figure}[p]
$$
\epsfxsize=5cm
\epsfbox{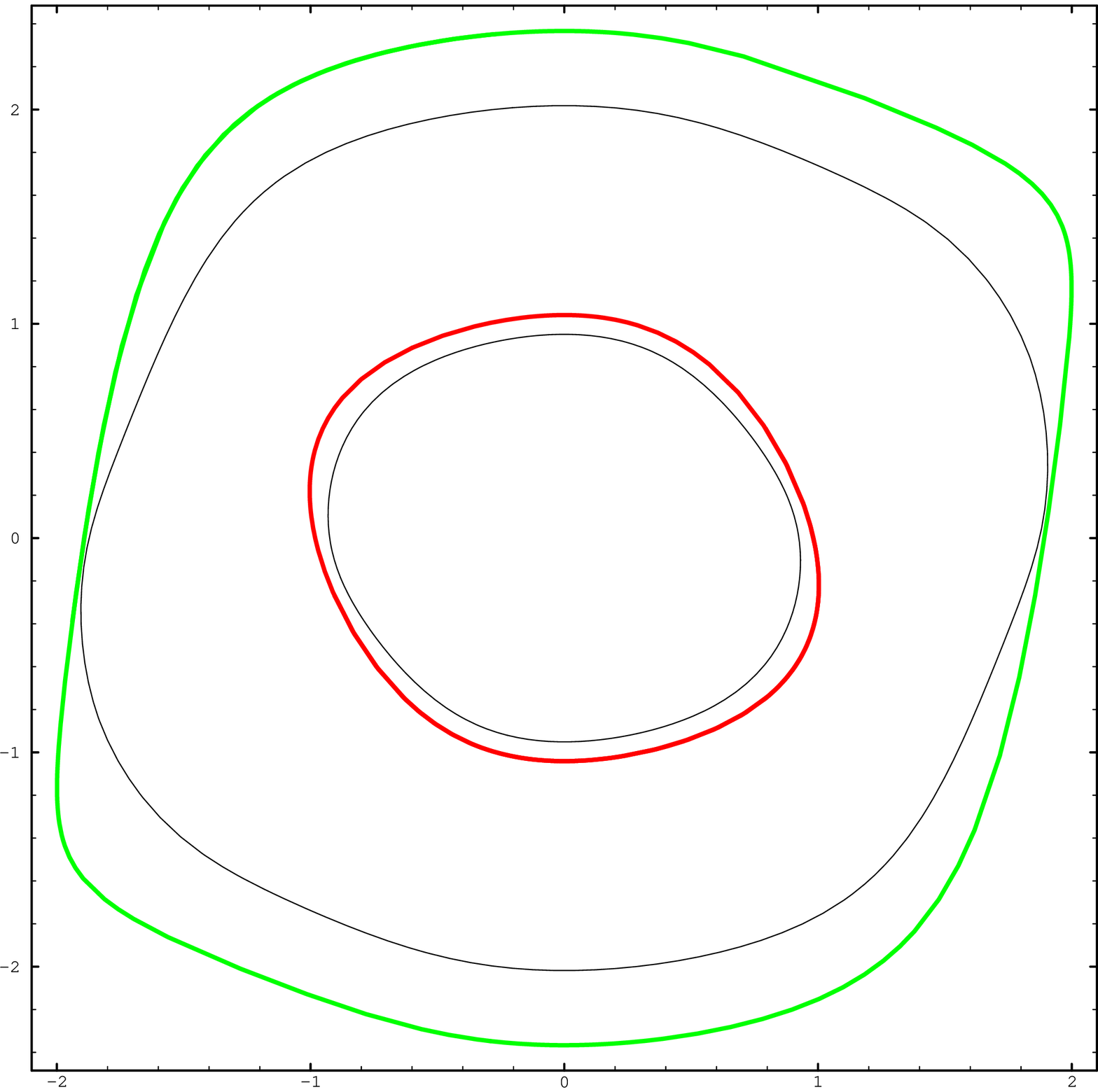}
$$
\caption{The limit cycles of equation (\ref{equa lienard ode bis}) with $F(x)$
given by (\ref{equa 2cy}) (rough curves) and their algebraic
 approximations (smooth curves)~: $h_6(x,y)=K_{6 \, 1} ^\star$ and
 $h_6(x,y)=K_{6 \, 2} ^\star$}
\label{fig 2cy 6}
\end{figure}

\begin{figure}[p]
$$
\epsfxsize=5cm
\epsfbox{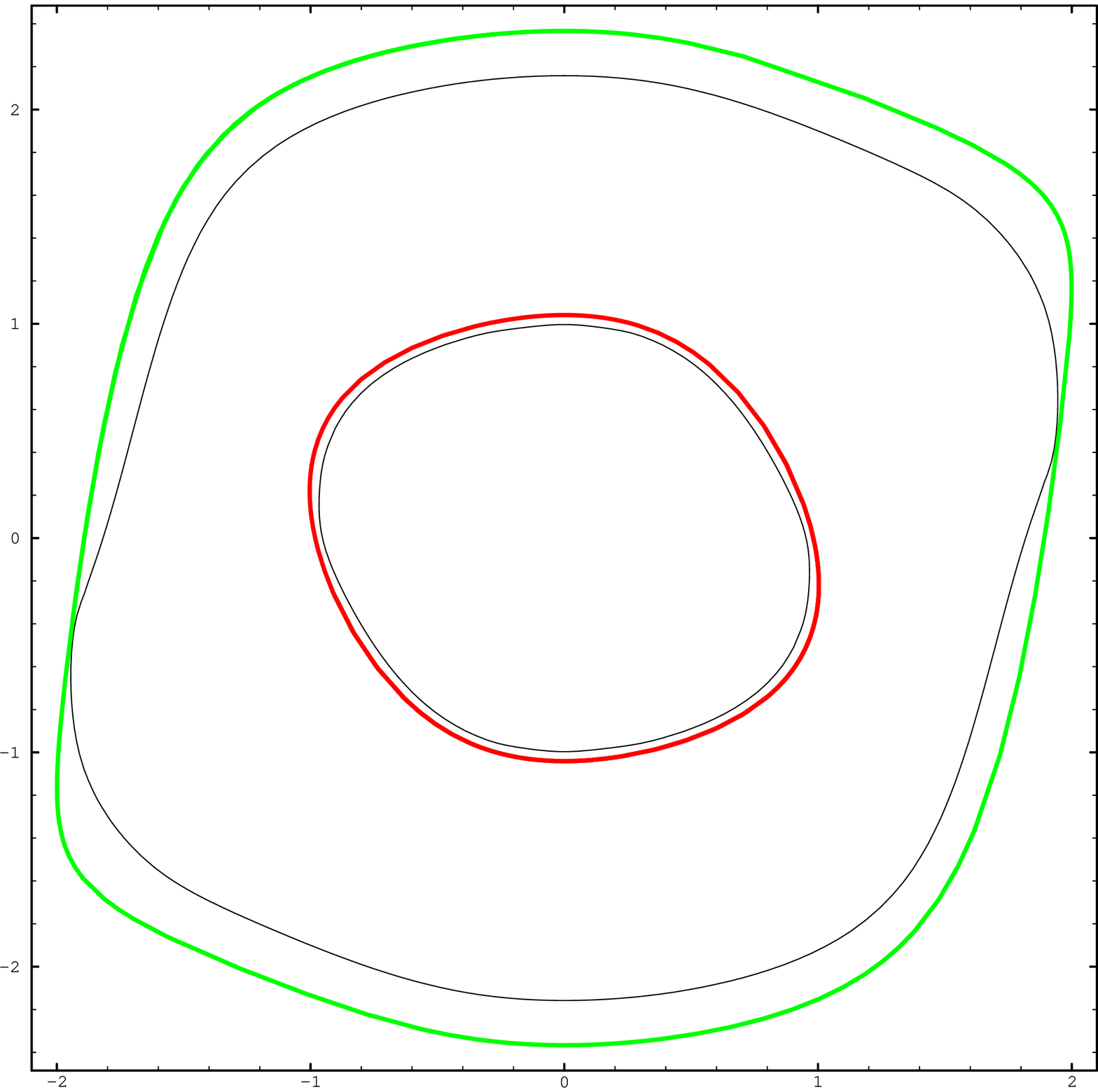}
$$
\caption{The limit cycles of equation (\ref{equa lienard ode bis})
 with $F(x)$
given by (\ref{equa 2cy}) (rough curves) and their algebraic
 approximations (smooth curves)~: $h_{14}(x,y)=K_{14 \, 1} ^\star$ and
 $h_{14}(x,y)=K_{14 \, 2} ^\star$}
\label{fig 2cy 14}
\end{figure}

 In fig. \ref{fig 2cy 6}
 and \ref{fig 2cy 14} we show these curves for $n=6$ and
 $n=14$, respectively. We also show the limit cycles obtained 
by numerical integration.
 In table \ref{table 2cy}, we give the
 values of the roots of $R_n(x)$ and the values of
 $K_{n1} ^\star$ and $ K_{n2} ^\star$ for $2 \leq n \leq 16$.
 These roots seem to
 converge to the maximum values of $x$ for each cycle
(the numerical values of the maximum of $x$ on each limit cycle
are $x_{max,1} \simeq 1.0034$ 
and $x_{max,2} \simeq 1.9992$ respectively).
 The curves $h_n(x,y)=K_{n1} ^\star $ and 
$h_n(x,y)= K_{n2} ^\star $ seem to converge to
 each one of the limit cycles of the system.\\

\begin{figure}[p]
 $$
 \epsfxsize=5cm
\epsfbox{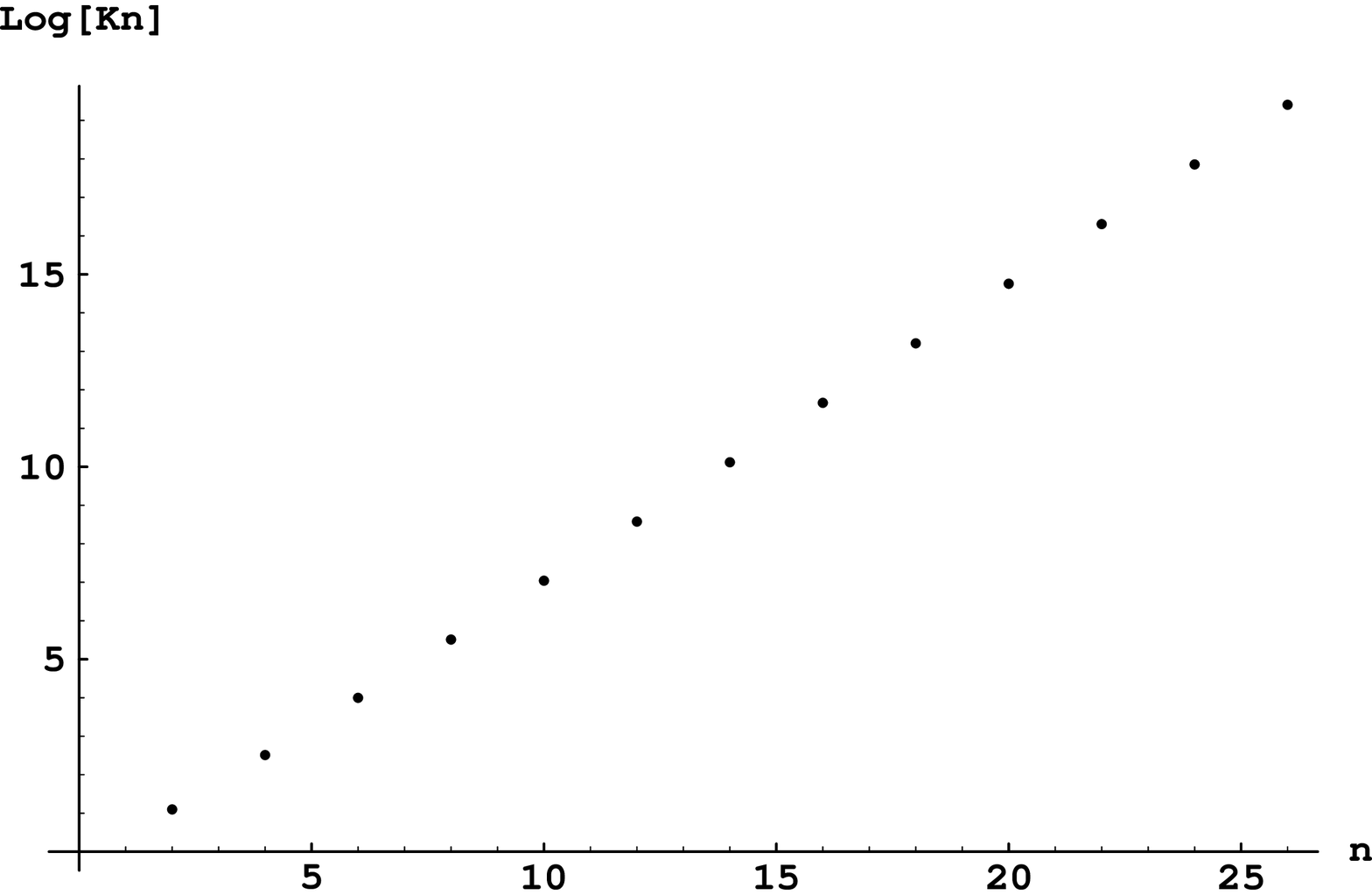}
$$
\caption{We show, for the van der Pol equation, the curve
$Log(K_n^\star)$ in function of $n$}
\label{logkn}
\end{figure}

For all the cases that we have studied, we have found that the 
values of the constants $K_n^\star$ go to zero or infinity
when $n \rightarrow \infty$.
In fact, it is easy to see from table \ref{table vdp} and 
table \ref{table 2cy} that the asymptotic behaviour of
 $K_n^\star$
with $n$ (for a given limit cycle), is given by
\begin{equation}
K_n^\star \simeq a (x_{max})^n
\end{equation}
where a is a constant which depends on the cycle 
(see fig. \ref{logkn}).\\

We have also considered system (\ref{equa lienard ode bis}) with~:
\begin{equation}
F(x)=x^5-\mu x^3+x  \label{rychkov mu}
\end{equation}
where $\mu$ is an arbitrary parameter.
It has been proved in \cite{rychkov} that this system has exactly two 
limit cycles for $\mu > 2.5$. It is clear that this system has
 no limit cycle
for $\mu <2$ because $r_2=0$ in that case. Hence, between $\mu=2$
 and $\mu=2.5$
there is a bifurcation value $\mu ^\star$ such that for
 $\mu < \mu ^\star$
the system has no limit cycles and for $\mu > \mu ^ \star$ the system has
 exactly two limit cycles. When $\mu=\mu ^\star$ the system undergoes a
 saddle-node bifurcation.

By applying our method, we can obtain lower bounds for the value of
 $\mu ^\star $. For each even value of $n$ we calculate the maximum 
value of $\mu$ for which $r_n$ is zero. This value of $\mu$ represents
 a lower bound for $\mu ^\star$. The results of
 these calculations are given
in table \ref{kov mu}.
The values of ${\mu}_n^\star$ seem to converge very quickly, in 
a monotonous way, when $n \rightarrow \infty$. Numerical integrations
of system (\ref{equa lienard ode bis}) with $F(x)$
 given by (\ref{rychkov mu})
seem to confirm that
 $\lim_{n \rightarrow \infty} \mu_n^\star=\mu^\star$.

Let us point out that it is the first time, in our knowledge,
that a bifurcation value of this type can be estimated in such
a way, that is by employing an analytical method instead of
a numerical integration of the system.\\

We have also analysed system (\ref{equa lienard ode bis}) with $F(x)$ 
given by~:
\begin{equation}
F(x)=x(x^2-1.6^2)(x^2-4)(x^2-9)
\end{equation}
For this case we have $r_2=r_4=3$. However, the second positive root
 of $R_4(x)$
is smaller than the second positive root of $R_2(x)$. Indeed for $n=6$
 we find $r_6=1$.
An annihilation of two roots has occured and this
 phenomenon has been annonced by the lowering of the value of one of the
 roots of $R_n(x)$. We conjecture that $r_n=1 \; \forall \, n$ even, greater than 4. The
 numerical analysis of this system seems to indicate that it has exactly 
one limit cycle.\\

For all the cases that we have studied, we have found that
two types of behaviour of $r_n$ are possible~:
\begin{itemize}
\item [{\bf i }] $r_n=r_n'$ for arbitrary even values of $n$ and $n'$. In 
this case
the number of limit cycles of the system is given by this common value
 of the number of positive roots of odd multiplicity of $R_n(x)$.
\item[{\bf ii }] the values of $r_n$ changes with $n$; in this case the
 values of $r_n$ decreases with $n$; moreover we have $r_n-r_n'=2p$ for
 $n'>n$ and $p \in {\Bbb N}$. The roots of $R_n(x)$ seem to disappear
 by pairs, when $n$ increases.
\end{itemize}
Guided by the particular cases that we have analysed, we
 establish the following conjecture~:\\ {\bf Conjecture~:}
{\it  Let be $l$ the number of limit cycles of
 (\ref{equa lienard ode bis}). Let be $r_n$
the number of positive roots of $R_n(x)$ (with $n$ even) of odd multiplicity.
Then we have~:
\begin{itemize}
\item [\bf{i }] $l \leq r_n \; \forall \, n$ even
\item [\bf{ii }] if $n' > n$ then $r_n-r_n'=2p$ with $p \in {\Bbb N}$.
\end{itemize}}

We have also analysed the roots of the polynomials $g_{j,n}$,
with $0 \leq j \leq n-1$. For odd values of $j$, the
roots of these polynomials are also related to the number and location
of the limit cycles of the system. For instance, for the van
der Pol equation, the polynomials $g_{j,n}(x)$ with $j$ odd have
exactly one positive root of odd multiplicity. These roots are an upper
bound to $x_{max}$. For a given odd value of $j$, the sequence
of roots of $g_{j,n}(x)$ decreases monotonously with $n$ and seems to
converge to the value of $x_{max}$. The best upper bounds are given
by the roots of $g_{1,n}(x)$, as can be seen in table \ref{table g}.
The reasons of such a behaviour of the roots of the polynomials 
$g_{j,n}(x)$ with $j$ odd are not clear to us.\\

We have shown in this paper that the polynomials 
$h_n(x,y)=y^n+g_{n-1,n}(x) y^{n-1}+g_{n-2,n}(x) y^{n-2}
+...+g_{1,n}(x) y + g_{0,n}(x)$
give a lot of information about the number and location of
 the limit cycles of (\ref{equa lienard ode bis}), in the case 
where $F(x)$ is an odd polynomial (for the case where $F(x)$ is
 not an odd polynomial, the limit cycles are not invariant
  under the transformation $(x,y) \rightarrow (-x,-y)$ and the results are
 not conclusive).
 The curves $ h_n(x,y)=K_n ^\star $
give algebraic approximations to each limit cycle
. These algebraic approximations seem to
converge to the limit cycles of the system. The positive roots
of odd multiplicity of
the polynomials $R_n(x)=\dot{h_n}(x,y)$ are related to the number of
limit cycles of (\ref{equa lienard ode bis}) and they give lower bounds
for the values of $x_{max}$ of each limit cycle. Moreover, the
roots of $g_{j,n}(x)$, with odd values of j, are also related
to the number of limit cycles and they give upper bounds to the
 value of $x_{max}$ for each limit cycle.

All the relevant information about the limit cycles of
 (\ref{equa lienard ode bis}) seems to be contained
in the polynomials $h_n(x,y)$. These polynomials are very easy to 
calculate with an algebraic manipulator program.

\clearpage
\begin{table}[t]
\begin{center}
\begin{tabular}{|c||c|c|c|c|c|c|c|c|c|c|}
\hline
n & 2 & 4 & 6 & 8 & 10 & 12 & 14 & 16 & 18 & 20 \\
\hline
root & 1.732 & 1.824 & 1.869 & 1.896 & 1.914 & 1.927 & 1.937
 & 1.944 & 1.950 & 1.955 \\
\hline
$K_n ^\star$ & 3 & 12.3 & 54.5 & 247.6 & 1141 & 5305 & 24773
 & 116050 & 544800 & $ \sim 2 \cdot  10^6 $ \\
\hline
\end{tabular}
\end{center}
\caption{For each value of $n$ we give the value of the root
 of $R_n(x)$ and the value of $K_n ^\star $ for the van der
 Pol equation.}
\label{table vdp}
\end{table}

\begin{table}[b]
\begin{center}
\begin{tabular}{|c|c|c|c|c|}
\hline
n & root one & $K_{n1}^\star $ & root two & $K_{n2} ^ \star$ \\
\hline
2 & 0.852 & 0.726 & 1.854 & 3.439 \\
4 & 0.905 & 0.711 & 1.885 & 14.5 \\
6 & 0.931 & 0.739 & 1.905 & 67.59 \\
8 & 0.945 & 0.784 & 1.920 & 334 \\
10 & 0.955 & 0.840 & 1.931 & 1712 \\
12 & 0.962 & 0.903 & 1.938 & 8973 \\
14 & 0.967 & 0.974 & 1.945 & 47741 \\
16 & 0.971 & 1.052 & 1.950 & 254400 \\
\hline
\end{tabular}
\end{center}
\caption{For each value of $n$, we give the two roots of $R_n(x)$
 and the values of $K_{n \, 1} ^\star$ and $K_{n \, 2} ^\star$
for equations (\ref{equa lienard ode bis}), with $F(x)$ given by (\ref{equa 2cy})}
\label{table 2cy}
\end{table}

\clearpage

\begin{table}[t]
\begin{center}
\begin{tabular}{|c||c|c|c|c|c|c|c|c|c|c|}
\hline
n &  2 & 4 & 6 & 8 & 10 & 12 & 14 & 16 & 18 & 20 \\
\hline
$\mu_n ^\star$  &  2 & 2.057 & 2.079 & 2.090 & 2.096 &
 2.100 & 2.103 & 2.105 & 2.106 & 2.107 \\
\hline
\end{tabular}
\end{center}
\caption{We give in this table, for each even value of $n$ between 2 and 20, 
a lower bound $\mu _n ^\star $ for the value of $\mu ^\star$. This sequence
 seems to converge rapidly toward $\mu ^\star $.}
\label{kov mu}
\end{table}
\begin{table}[b]
\begin{center}
\begin{tabular}{|c|c|c|c|}
\hline
n      &   Root of $ R_n$  & 
 Root of $g_{1,n}$ & Root of $g_{3,n}$ \\
\hline
2 & 		1.7321    &     ---     &       --- \\
4  &            1.8248    &     2.2361 &        --- \\
6   &           1.8697     &    2.1924  &       2.2361 \\
8    &          1.8965      &   2.1658   &      2.2063 \\
10    &         1.9144   &      2.1475    &     2.1854 \\
12     &        1.9273    &     2.1341     &    2.1697 \\
14      &       1.937      &    2.1236      &   2.1574 \\
16       &      1.9446      &   2.1152       &  2.1474 \\
18        &     1.9507       &  2.1083    &     2.1391  \\
20         &    1.9558        & 2.1025     &    2.1321 \\
\hline
\end{tabular}
\end{center}
\caption{For each even value of $n$, between 2 and 20, 
we give the roots of the polynomials $R_n$, $g_{1,n}$
 and $g_{3,n}$ respectively, for
the van der Pol equation.}
\label{table g}
\end{table}

\clearpage


\clearpage

\begin{thebibliography}{15}

\bibitem{strogatz} S. H. Strogatz, {\it Nonlinear Dynamics and Chaos}
(Addison Wesley, 1994).

\bibitem{nayfeh} A. Nayfeh and B. Balachandran, 
{\it Applied Nonlinear Dynamics}
(J. Wiley, New York, 1995).

\bibitem{kerhet}L. Edelstein-Keshet, {\it Mathematical Models in Biology}
(Random House, New York, 1988).

\bibitem{salamich2} L. Salasnich, ``Instabilities, Point Attractors
and Limit Cycles in a Inflationary Universe'',
Mod. Phys. Lett. {\bf A10}, 3119 (1995).

\bibitem{poland} D. Poland, ``Loci of limit cycles'',
 Phys. Rev. E {\bf 49}, 157 (1994).

\bibitem{salamich} L. Salasnich, ``On the Limit Cycle of an 
Inflationary Universe'', Preprint Universita di Padova (1996).

\bibitem{qian} Ye Yan-Qian et al., {\it Theory of Limit Cycles,
translations of mathematical monographs}, Vol. 66
(American Mathematical Society, Providence, 1986).

\bibitem{maier} R. Maier and D. Stein,
``Oscillatory behaviour of the rate of escape through
an unstable limit cycle'', Phys. Rev. Lett. {\bf 77}, 
4860 (1996).

\bibitem{farkas} M. Farkas, {\it Periodic Motion} (Springer, Berlin, 1994).

\bibitem{viano} H.Giacomini and M.Viano, ``Determination of limit
cycles for two-dimensional dynamical systems'', Physical 
Review E {\bf 52}, 222 (1995).

\bibitem{llibre} H. Giacomini, J. Llibre and M. Viano,
 ``On the nonexistence, existence and uniqueness of limit
cycles'', Nonlinearity {\bf 9}, 501 (1996).

\bibitem{delamotte} B. Delamotte,
 ``A non pertubative method for solving differential
 equations and finding limit cycles'', Phys. Rev. 
Lett. {\bf 70}, 3361 (1993).

\bibitem{rayleigh} J. Rayleigh, {\it The Theory of Sound}
, New York, Dover (1945).

\bibitem{van der pol} B. van der Pol, ``Forced oscillations
in a circuit with nonlinear resistance'', London, Edinburgh
 and Dublin Phil. Mag. {\bf 3}, 65 (1927).

\bibitem{lienard} A. Li\'enard, ``Etude des oscillations entretenues'',
 Rev. Gen. d'\'electricit\'e, 
XXIII, 901 (1928).

\bibitem{smith} N. Levinson and D. Smith, ``A genereal equation for
relaxation oscillations'', Duke Math. 
Journal {\bf 9}, 382 (1942).

\bibitem{de melo} A. Lins, W. de Melo and C. Pugh,
 {\it On Li\'enard's equation}, Lecture Notes in Mathematics
 {\bf 597}, 335, Springer-Verlag (1977).

\bibitem{perko} L. Perko,
{\it Differential Equations and Dynamical Systems}
 (Springer-Verlag, second edition, 1996).

\bibitem{lloyd} N.G. Lloyd, New Directions in Dynamical
Systems, London Math. Soc. Lecture Note Series N.127, edited
by T.Bedford and J. Swift (Cambridge University Press, 1988).

\bibitem{dumortier2} F. Dumortier and C. Rousseau,
``Cubic Li\'enard equations with linear damping'',
 Nonlinearity {\bf 3}, 1015 (1990).

\bibitem{dumortier} F. Dumortier and C. Li,
``On the uniqueness of limit cycles surrounding 
one or more singularities for Li\'enard equations'',
 Nonlinearity {\bf 9}, 1489 (1996).

\bibitem{blows and lloyd} T.Blows and N.Lloyd,
``The number of small amplitude limit cycles of
 Li\'enard equations'', Math. Proc. Cambridge Phil.
 Soc. {\bf 95} (1984), 751.

\bibitem{xun} Xun-Cheng Huang, ``Uniqueness of limit cycles of
generalized Li\'enard systems and predator-prey systems'', 
J. Phys. A~: Math. Gen. {\bf 21}, L685  (1988).

\bibitem{yang} Yang Kuang and H. Freedman, ``Uniqueness of limit cycles
in Gause-type models of predator-prey systems'', Math. Biosciences {\bf 88},
67 (1988).

\bibitem{rychkov} G.S. Rychkov, ``The maximal number of
 limit cycles of the system
 $\dot{y}=-x \, \mbox{, } \dot{x}=y-(a_1x+a_3x^3+a_5x^5)$
 is equal to two'', Differential Equations {\bf 11},
(1975), 301.

\end{thebibliography}
\end{document}